\documentclass[aps,superscriptaddress,showpacs,prl,twocolumn,longbibliography]{revtex4-1}

\usepackage[pdftex]{graphicx}
\usepackage{amssymb}
\usepackage{amsmath}
\usepackage{xspace}
\usepackage{color}
\usepackage{hyperref}
\usepackage{bm}
\usepackage{mathtools}

\begin{document} 

\title{Charge-order melting in the one-dimensional Edwards model}

\author{Florian Lange}
\affiliation{Friedrich-Alexander-Universit\"at Erlangen-N\"urnberg (FAU), Erlangen National High Performance Computing Center (NHR@FAU), 91058 Erlangen, Germany}
\author{Gerhard Wellein}
\affiliation{Friedrich-Alexander-Universit\"at Erlangen-N\"urnberg (FAU), Erlangen National High Performance Computing Center (NHR@FAU), 91058 Erlangen, Germany}
\author{Holger Fehske}
\affiliation{Friedrich-Alexander-Universit\"at Erlangen-N\"urnberg (FAU), Erlangen National High Performance Computing Center (NHR@FAU), 91058 Erlangen, Germany}
\affiliation{Institute of Physics, University of Greifswald, 17489 Greifswald, Germany}

\date{\today}

\begin{abstract}
  We use infinite matrix-product-state techniques to study the time evolution of the charge-density-wave (CDW) order
  after a quench or a light pulse in a fundamental fermion-boson model. 
  The motion of fermions in the model is linked to the creation of bosonic excitations, which counteracts the melting of the CDW order. 
  For low-energy quenches corresponding to a change of the boson relaxation rate, we find behavior similar to that in an effective $t$-$V$ model. 
  When the boson energy is quenched instead or a light pulse is applied to the system, the transient dynamics are more complex,
  with the CDW order first quickly decreasing to an intermediate value while the density-wave-like order of the bosons rises.  
  In the case of pulse irradiation, the subsequent time-evolution of the CDW order depends strongly on the photon frequency. For frequencies slightly below the boson energy, we observe a temporary increase of the CDW order parameter. Our results reveal the complex physics of driven Mott insulators in low-dimensional systems with strong correlations.  
\end{abstract}

\maketitle

\paragraph{Introduction.}
Ultrafast spectroscopy~\cite{UltrafastSpectroscopy} and experiments on ultracold atoms in optical lattices~\cite{OpticalLattices} have allowed to directly observe the dynamics of quantum states in strongly-correlated materials. 
From a theoretical perspective, such non-equilibrium systems
are intriguing, as they can give rise to exotic metastable states~\cite{PhysRevLett.101.265301,EtaPairingKaneko}, and provide insight into the relaxation dynamics of quantum matter~\cite{PhysRevLett.105.257001,Ueda2020}. 
For materials with long-ranged order, a natural question is how the order-parameter evolves after a perturbation, e.g., a quench or a light pulse~\cite{PhysRevLett.105.187401,Maklar2021}. In the context of one-dimensional (1D) systems, this has been addressed by numerical simulations for the magnetic order in spin chains~\cite{RelaxationAFMOrder,HowOrderMelts} as well as for charge density waves (CDWs) of electrons~\cite{MeltingHolsteinPulse,MeltingHolsteinQuench}. For the latter, it was found that an electron-phonon coupling modeled by a Holstein Hamiltonian not only renormalizes the electron mass via the formation of polarons, but can significantly alter the transient dynamics. 
This highlights that to faithfully model a system's dynamics, it is important to take into account possible couplings to environmental degrees of freedom.

The paradigmatic Edwards model describes particles whose movement requires the creation or annihilation of local bosons that parametrize the interaction with a background medium~\cite{EdwardsFirst}. 
This is a very generic situation in a great variety of condensed matter systems, in which the background can be seen as a deformable lattice (phonons), spins or orbitals forming, e.g., an ordered structure~\cite{Berciu2009,Alex07,PhysRevB.39.6880,PhysRevB.79.224433}.  Previous studies have focused on transport properties~\cite{EdwardsAlv1,EdwardsAlv2} and the ground-state phase diagram~\cite{EdwardsFirst,EdwardsPRL101,EdwardsPRL102}. At half band-filling, it was demonstrated in particular that there is a metal-insulator transition between a Tomonaga Luttinger liquid (TLL) and a CDW phase in one dimension. However, despite the rich physics captured by the Edwards model, its non-equilibrium properties are completely unexplored so far.  

In this work, we utilize infinite matrix-product-state (MPS) methods to study the melting of the CDW order in the 1D Edwards model and discuss 
the differences compared with the previously investigated $t$-$V$ and Holstein models. We consider both sudden quenches of the Hamiltonian parameters and pulse irradiation, which could be relevant for pump-probe experiments.

\paragraph{Model and method.}
The Hamiltonian of the 1D Edwards model is 
\begin{align}
  \hat{H} &= -t_b \sum_{i} \left( \hat{f}_{i+1}^\dagger \hat{f}_{i}^{\phantom{\dagger}} ( \hat{b}_{i}^\dagger + \hat{b}_{i+1}^{\phantom{\dagger}}) + \text{H.c.} \right) \nonumber \\ & \hspace*{1.5cm}
   - \lambda \sum_i (\hat{b}_i^\dagger + \hat{b}_i^{\phantom{\dagger}}) + \omega_0 \sum_i \hat{b}_i^\dagger \hat{b}_i^{\phantom{\dagger}} ,
  \label{eqHam}
\end{align}
where $\hat{f}^{(\dagger)}$ and $\hat{b}^{(\dagger)}$ are fermion and boson annihilation (creation) operators, respectively.  Therefore $\hat{H}$ describes boson-affected quantum transport: A fermion emits or absorbs a local boson of energy  $\omega_0$ every time it hops between neighboring lattice sites. The bosons can relax via the $\lambda$-term. Obviously, large $\omega_0$ and small $\lambda$ parametrize a rather stiff background medium. 
A different representation of the model is obtained by the unitary transformation $e^{\sum_i(\hat{b}_i - \hat{b}_i^{\dagger}) \lambda / \omega_0}$, which removes the $\lambda$-term but adds a free nearest-neighbor hopping $-t_f \sum_{i} ( \hat{f}_{i+1}^\dagger \hat{f}_{i}^{\phantom{\dagger}} + \text{H.c.} ) $ with a renormalized transfer integral $t_f = 2 \lambda t_b / \omega_0$.

\begin{figure}[tb]
  \centering
  \includegraphics[width=0.8\linewidth]{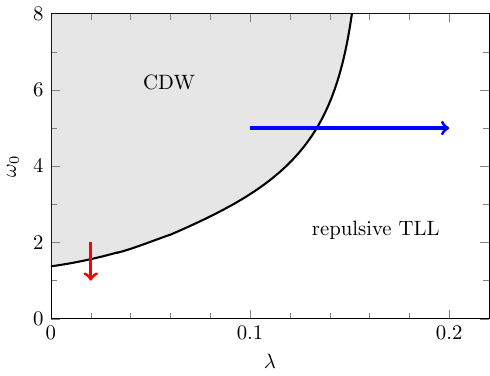}
  \caption{Ground-state phase diagram of the 1D Edwards model at half-filling according to Ref.~\cite{EdwardsPRL102}. The red and blue arrows indicate the quenches studied in Figs.~\ref{fig_quench1}(a) and \ref{fig_quench2} below, respectively.} 
  \label{fig_pd}
\end{figure}

We assume a half-filled electron band, in which case the ground state of the Edwards model is either a TLL or a Mott insulator with CDW order~\cite{EdwardsPRL101}.  The phase diagram was derived in Ref.~\cite{EdwardsPRL102} by  density-matrix-renormalization group calculations of the TLL parameter and the charge gap. For the benefit of the reader it is redrawn in Fig.~\ref{fig_pd}. The CDW phase is characterized by a finite value of the electronic order parameter
\begin{align}
  \mathcal{O}_{\text{CDW}} &= \frac{2}{N} \sum_i (-1)^i  \langle  \hat{f}_i^\dagger \hat{f}_i^{\phantom{\dagger}} \rangle ,
\end{align}
where $N$ is the number of sites. Since the broken translation symmetry in the CDW phase is reflected in the boson densities as well, it makes sense to also consider the bosonic order parameter 
\begin{align}
  \mathcal{O}_{b} = \frac{2}{N}  \sum_i (-1)^i  \langle  \hat{b}_i^\dagger \hat{b}_i^{\phantom{\dagger}} \rangle .
\end{align}
Importantly, $\mathcal{O}_{\text{CDW}} + 2 \mathcal{O}_{b}$ is conserved in the limit $\lambda = 0$.

To simulate the time evolution after the system is driven out of equilibrium, we employ numerical methods based on infinite MPS. 
Namely, we first calculate the ground state with the variational method of Ref.~\cite{VUMPS}, and then carry out the time evolution using the infinite time-evolving block decimation (iTEBD)~\cite{iTEBD} with a second-order Suzuki-Trotter decomposition and time step $0.025\,t_b^{-1}$. The fermion and boson degrees of freedom are treated as separate sites in the MPS to keep the dimension of the local Hilbert spaces small. Since this increases the range of the couplings, we use the method of Ref.~\cite{iTEBDHybrid} and apply the iTEBD gates approximately using the time-dependent variational principle (TDVP)~\cite{TDVPforMPS}. 

The boson Hilbert spaces need to be truncated to a finite dimension $D_b$. Because the Mott insulator in the Edwards model typifies a CDW phase with low boson density, a relatively small value of $D_b$ is sufficient and methods to treat large local Hilbert spaces~\cite{PhysRevB.57.6376,PhysRevB.92.241106,10.21468/SciPostPhys.10.3.058,LargeLocalSpaceComparative} are not necessary. We use $D_b=6$ for the simulations in this work, and take $t_b = 1$ as the unit of energy. 
The truncation error in the two-site TDVP is kept below $5 \cdot 10^{-7}$, which leads to a maxmimum MPS bond dimension of $3600$. We have confirmed that the results are converged by doing additional simulations with lower bond dimensions.

\begin{figure}[tb]
  \centering
  \includegraphics[width=0.9\linewidth]{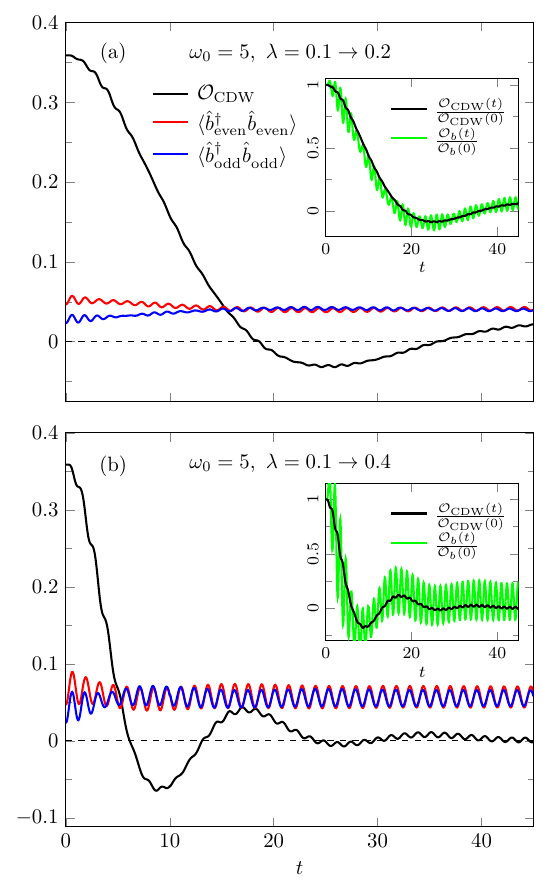}
  \caption{Time evolution of the CDW order parameter and boson densities for a quench from the CDW to the TLL phase at $\omega_0 = 5$. The initial state is the ground state for $\lambda = 0.1$, while the time evolution is according to the Hamiltonian with $\lambda = 0.2$ [panel (a)] or $\lambda = 0.4$ [panel (b)]. 
    $\hat{b}_{\text{even}}$ ($\hat{b}_{\text{odd}}$) is a boson annihilation operator at an arbitrary even (odd) site.}
  \label{fig_quench1}
\end{figure}

\paragraph{Quenches.}
The simplest setting to study the melting of the CDW order is a quench of one of the Hamiltonian parameters. 
Here, we restrict ourselves to quenches from the CDW to the TLL phase [see Fig.~\ref{fig_pd}], and assume that the system is initially in the ground state. Figure~\ref{fig_quench1}(a) shows the time evolution of the order parameter $\mathcal{O}_{\text{CDW}}$ and the boson densities after the boson-relaxation parameter is abruptly switched from $\lambda = 0.1$ to $ 0.2$ at fixed boson energy $\omega_0  = 5$.
The order parameter decreases and exhibits damped oscillations around zero. Similar behavior has been previously observed for quenches in the spin-1/2 XXZ chain and the 1D Holstein model~\cite{RelaxationAFMOrder,MeltingHolsteinQuench}.
As the boson energy $\omega_0 = 5$ is quite large and not modified by the quench, the boson densities do not change significantly during the time evolution.
The long-lived oscillations with angular frequency ${\simeq} \, \omega_0$ already appear for isolated boson sites and can be attributed to the change of the boson eigenmodes. There is also a small staggered order $\mathcal{O}_b$ of the bosons that, when appropriately scaled, closely follows that of the fermions. 
If $\lambda$ is quenched to $0.4$ instead [see Fig.~\ref{fig_quench1}(b)], the results are qualitatively similar, but $\mathcal{O}_{\text{CDW}}$ decays more quickly and the period of the oscillations is shortened. The oscillations of the boson densities also become more pronounced.

Overall, the above observations fit into a perturbative picture, in which the dynamics of the electrons are described by an effective $t$-$V$ model (see Supplemental Material~\cite{supplemental}) 
  and the stiff background contains almost no bosons. Because of the reduced energy scale of the effective model, the relaxation of the CDW order is slower than in a model without boson coupling and similar hopping amplitude, however.  

\begin{figure}[tb]
  \centering
  \includegraphics[width=0.9\linewidth]{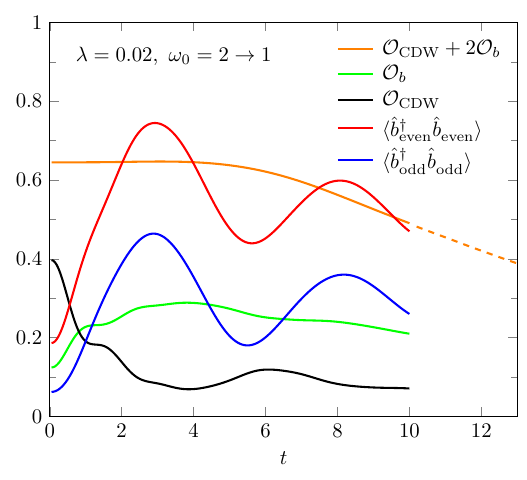}
  \caption{CDW order parameter and boson densities for a quench from the CDW to the TLL phase at $\lambda = 0.02$. The boson energies before and after the quench are $\omega_0 = 2$ and $\omega_0 = 1$, respectively. }
  \label{fig_quench2}
\end{figure}

As a second variant,  we consider the quench from $\omega_0  = 2$ to $1$ at constant $\lambda = 0.02$. Here, the perturbative model should not be applicable, since $\omega_0$ and $t_b$ are of the same order. 
Another difference is that the energy density relative to the ground state is much larger that in the previous case. As shown in Fig.~\ref{fig_quench2}(b), the CDW order $\mathcal{O}_{\text{CDW}}$ again decreases after the quench. 
However, although the Hamiltonian parameters for the time-evolution 
correspond to a point in the TLL phase, this is accompanied by an increase of the boson order $\mathcal{O}_{b}$. 
In fact, $\mathcal{O}_{\text{CDW}} + 2 \mathcal{O}_{b}$ remains approximately constant until time $t \simeq 4$, and only slowly decreases afterwards. This shows that the initial dynamics can be mainly attributed to the boson-affected hopping ${\propto} \, t_b$, while the free hopping term ${\propto} \, t_f$ in the alternate representation of the model becomes important for the long-time behavior. 
Both the fermion and boson orders appear to decay. 
Simulating the long-time evolution with MPS to determine the relaxation times is not feasible, however, because we can only reach times about $t = 10$ before the required bond dimension becomes prohibitively large. 

We have attempted to push the simulation to slightly longer times by shifting a part of the time evolution to the operators that need to be evaluated~\cite{tDMRGHeisenbergPicture,KENNES201637}. 
In this approach, the operators are expressed as matrix-product operators (MPO) and evolved according to the Heisenberg picture until their bond dimension becomes too large. Unfortunately, we found that in the Edwards model the growth of the bond dimension 
is too fast to simulate significantly longer times. 
The quantity $\mathcal{O}_{\text{CDW}} + 2 \mathcal{O}_{b}$ can be evolved slightly longer, since it is conserved in the limit $\lambda = 0$.
To take advantage of the small $\lambda$, we express the operator in momentum space by adapting the method of Ref.~\cite{PhysRevB.105.205130} to MPOs. The result of combining the MPS and MPO simulations is displayed in Fig.~\ref{fig_quench2} as a dashed line. It is consistent with
a continued slow decay of the density-wave order. 

\paragraph{Pulse irradiation.}

\begin{figure}[tb]
  \centering
  \includegraphics[width=0.95\linewidth]{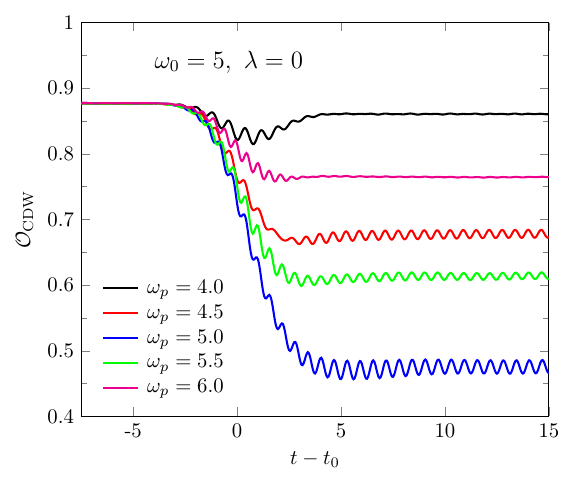}
  \caption{Time evolution of the CDW order parameter for pulses with different frequencies. Model parameters are $\omega_0 = 5$ and $\lambda = 0$.}
  \label{fig_pulse1}
\end{figure}

A different way to melt the CDW order, which is motivated by pump-probe experiments, is to apply a light pulse to the system. 
The corresponding time-dependent Hamiltonian is obtained by a Peierls substitution
$\hat{f}_{j+1}^\dagger \hat{f}_j^{\phantom{\dagger}} ( \hat{b}_j^\dagger + \hat{b}_{j+1}^{\phantom{\dagger}}) \to e^{-i A(t)} \hat{f}_{j+1}^\dagger \hat{f}_j^{\phantom{\dagger}} ( \hat{b}_j^\dagger + \hat{b}_{j+1}^{\phantom{\dagger}})$
in the hopping term. We consider a Gaussian pulse centered around $t_0$ with width $\sigma_p$, amplitude $A_0$ and frequency $\omega_p$, i.e., the vector potential has the form
\begin{align}
  A(t) = A_0 e^{-\frac{(t-t_0)}{2 \sigma_p^2}} \cos[\omega_p(t-t_0)] .
\end{align}
In the following, $\sigma_p = 2$ and $A_0 = 0.2$ if not stated differently.  

Based on results for the optical response function~\cite{PhysRevB.80.155101}, we expect that the pulse has the strongest effect when its frequency $\omega_p$ is close to  the boson energy $\omega_0$. For finite $\lambda$, there should also be a noticeable response for smaller $\omega_p$.  
If $\omega_0$ is large, however, the physics in that case will be qualitatively similar to that in the 
$t$-$V$ model, for which the CDW melting due to a pulse has been studied previously~\cite{MeltingHolsteinPulse}. We therefore focus on frequencies 
$\omega_p \approx \omega_0$.

Figure~\ref{fig_pulse1} shows the time evolution of the CDW order after a pulse is applied to a system with $\omega_0 = 5$ and $\lambda = 0$.
While the order parameter quickly decreases during the pulse, it stays nearly constant afterwards, except for small, persistent oscillations. 
These oscillations have frequency $2 \omega_0$ and an amplitude scaling as $A_0^2$ for small pulse strengths, indicating that they can be attributed to two-boson excitations created by second-order transitions. 
That the CDW order does not melt completely is explained by the fact that for $\lambda = 0$, flipping the CDW order requires a large energy of approximately $ \omega_0 $ per unit cell. 
Moreover, since $\mathcal{O}_{\text{CDW}} + 2 \mathcal{O}_{b}$ is conserved, the order parameter for either fermions or bosons must remain finite in the long-time limit. The effect of the pulse is thus to decrease $\mathcal{O}_{\text{CDW}}$ and increase $\mathcal{O}_{b}$. It should be noted that the fermions can move by an even number of sites via effective higher-order hopping processes, so that the absence of CDW melting for $\lambda = 0$ does not apply to the CDW phase of the Edwards model at one-third filling~\cite{doi:10.7566/JPSCP.3.013006}. When $\omega_p$ is detuned relative to $\omega_0$, the results are similar but the reduction of the CDW order and the strength of the oscillations become weaker.

\begin{figure}[tb]
  \centering
  \includegraphics[width=0.97\linewidth]{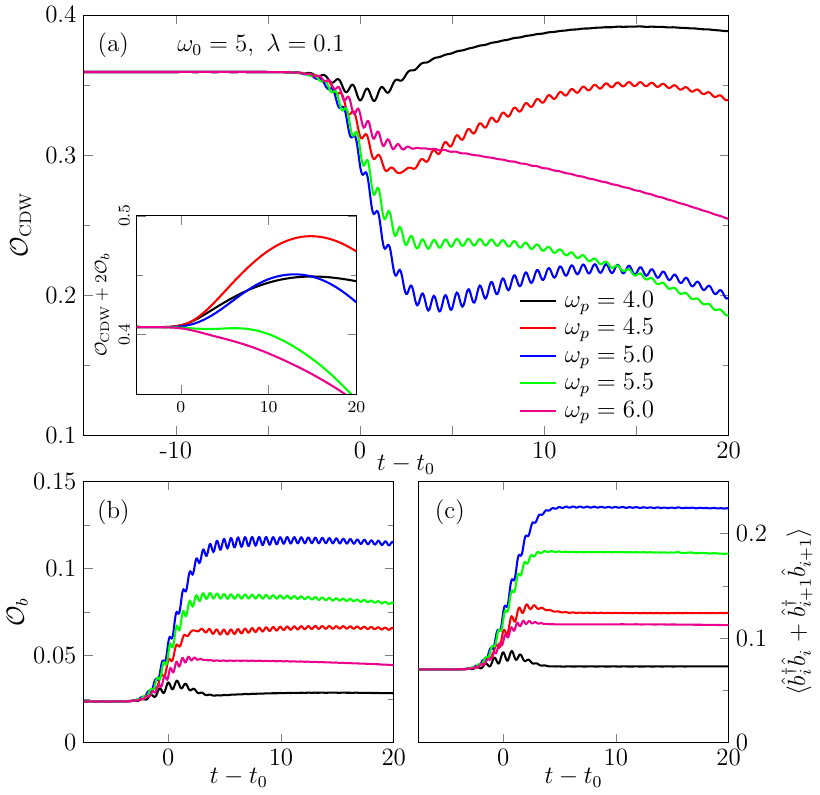}
  \caption{Time dependence of observables after a light pulse for model parameters $\omega_0 = 5$ and $\lambda = 0.1$. }
  \label{fig_pulse2}
\end{figure}

\begin{figure}[tb]
  \centering
  \includegraphics[width=0.97\linewidth]{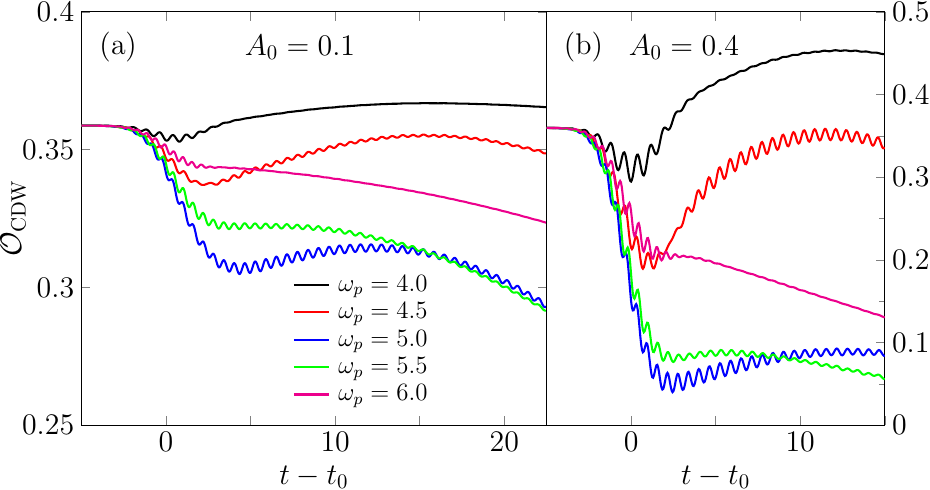}
  \caption{CDW order parameter after a light pulse with amplitude $A_0=0.1$ [panel (a)] or $A_0=0.4$ [panel (b)]. Other parameters are the same as in Fig.~\ref{fig_pulse2}.} 
  \label{fig_pulse3}
\end{figure}

More complex dynamics can be expected when there is a 
finite boson relaxation $\lambda = 0.1$ [see Fig.~\ref{fig_pulse2}]. 
Similarly as for $\lambda = 0$, $\mathcal{O}_{\text{CDW}}$ ($\mathcal{O}_{b}$) initially decreases (increases).
The CDW order now continues to change after the pulse, however, because of the effective free hopping with amplitude $t_f$. 
Interestingly, in this transient regime $\mathcal{O}_{\text{CDW}}$ increases for some pulse frequencies, and can even exceed the ground-state value before it falls off again. 
This time-dependence of the CDW order, which is fundamentally  different from the monotonic exponential decay found in Ref.~\cite{MeltingHolsteinPulse} for pulse irradiation of the 1D Holstein model, does not require fine tuning of the pulse parameters and also occurs, e.g., for different values of the amplitude $A_0$ as demonstrated in Fig.~\ref{fig_pulse3}.

  A qualitative picture for the enhancement of the CDW order is provided by
  an effective Floquet model
  that is valid for large $\omega_0$ and frequencies $\omega_p$ sufficiently detuned from $\omega_0$~\cite{supplemental}. 
  Like in the case without driving, the effective Hamiltonian is that of a modified $t$-$V$-model, but the nearest-neighbor repulsion is amplified for $\omega_p < \omega_0$ and diminished for $\omega_p > \omega_0$. This agrees with the result in Fig.~\ref{fig_pulse2} that $\mathcal{O}_{\text{CDW}}$ grows for low pulse frequencies. 
  The influence of the increased repulsion can also be seen in the time evolution of $\mathcal{O}_{\text{CDW}} + 2 \mathcal{O}_{b}$ [inset of Fig.~\ref{fig_pulse2} (a)], which 
  unlike $\mathcal{O}_{\text{CDW}}$ does not exhibit a sharp drop-off due to the resonant excitation of bosons.

In constrast to the charge order of the fermions, both the boson density $\langle \hat{b}_i^\dagger \hat{b}_i^{\phantom{\dagger}} + \hat{b}_{i+1}^\dagger \hat{b}_{i+1}^{\phantom{\dagger}} \rangle$ and the order parameter $\mathcal{O}_{b}$ remain nearly constant after the pulse. 
The approximate conservation of the boson density is expected because the boson energy $\omega_0 = 5$ is large compared to the energy scale of the electronic excitations, i.e., $\omega_0 \gg t_f, \,t_b^2/\omega_0$. 
To explain the slow evolution of $\mathcal{O}_{b}$ it is helpful to look at the effective Hamiltonian~\cite{supplemental}, which shows that up to second order in the hopping bosons can not move between odd and even sites (at higher order, terms involving nearest-neighbor boson hopping that scale as $ \lambda t_b^3 / \omega_0^3 $ appear). 
Accordingly, the time scale for the relaxation of the boson order $\mathcal{O}_{b}$ should be much longer than that for the relaxation of $\mathcal{O}_{\text{CDW}}$. 

The effective model furthermore suggests that it is energetically favourable for the fermion and boson orders to be aligned so that $\mathcal{O}_{\text{CDW}}$ and $\mathcal{O}_b$ have the same sign. Since $\mathcal{O}_{b}$ seems to persists for long times, we may speculate that the fermion order $\mathcal{O}_{\text{CDW}}$ will first settle around a reduced but finite value before it slowly approaches zero. Similar observations have been made for certain quenches in the spinless Holstein model, where the staggered displacement of phonons in the initial state can remain for long times~\cite{MeltingHolsteinQuench}. A difference in the Edwards model is that the boson order $\mathcal{O}_{b}$ is rather small in the CDW ground states.
Because of the boson-controlled hopping, it instead increases when the CDW order of the fermions is disturbed by a light pulse or a quench to lower boson energies $\omega_0$.

\paragraph{Conclusions.}
We numerically studied the charge-order melting in terms of the 1D Edwards model, which is a minimal description for electrons interacting with a  background medium. 
The quench parameters were chosen to drive a CDW-TLL insulator-metal transition as observed in many low-dimensional materials. In the Edwards model, this transition is of the Mott-Hubbard type rather than the Peierls type. 
By using the real-time simulation of infinite matrix-product states, we avoided boundary and finite-size effects in the simulations. 
Although the simulatable times are too short for the system to reach a true stationary state, the main effects of the fermion-boson coupling on the relaxation dynamics can be observed. 
For quenches that inject a large-enough energy into the system, there is a rapid  
growth of the boson density on a time scale $t_b^{-1}$, along with a reduction of the fermion CDW order $\mathcal{O}_{\text{CDW}}$ and an increase of the boson density-wave order $\mathcal{O}_b$.
Applying a light pulse tuned to the boson energy $\omega_0$ has a similar initial effect, but the dynamics that follow differ in the two cases. 
For the quench, we find a slow decrease of both $\mathcal{O}_{\text{CDW}}$ and $\mathcal{O}_b$, while for the pulse, where the final Hamiltonian is still in the CDW regime with large $\omega_0$, only $\mathcal{O}_{\text{CDW}}$ continues to change significantly whereas the boson order $\mathcal{O}_b$ stays nearly constant. 
Assuming that thermalization occurs, both density-wave orders should eventually disappear. However, because of the different time scales for the dynamics of the fermion and boson orders, the system may first approach a quasi-stationary state with finite $\mathcal{O}_{\text{CDW}}$. A generalized Gibbs ensemble description based on the perturbative Hamiltonian as in Ref.~\cite{Murakami2022} could perhaps give more insight into the properties of such a non-equilibrium state.

  Remarkably, a pulse with frequency below the boson energy can lead to an increased CDW order at intermediate times. 
  This is explained by an effective Floquet Hamiltonian that has an amplified nearest-neighbor interaction compared with the Hamiltonian from perturbation theory without driving. The enhancement of the CDW order also occurs for smaller $\omega_0$, however, where the effective model is no longer accurate.

For future studies, it would be interesting to investigate how the non-equilibrium charge dynamics differ in more elaborate models that take into account, e.g., a dispersion of the bosons. 
Understanding the effect of the fermion-boson coupling on the entanglement dynamics should also be valuable, in particular with regard to recent results on the quantum Fisher information in non-equilibrium systems~\cite{EntanglementDynamicsQFI}. 

\begin{acknowledgments}
The authors gratefully acknowledge the scientific support
and HPC resources provided by the Erlangen National High
Performance Computing Center (NHR@FAU) of the Friedrich-Alexander-Universität Erlangen-Nürnberg (FAU). NHR funding is provided by federal and Bavarian state
authorities. NHR@FAU hardware is partially funded by the
German Research Foundation (DFG) – 440719683.   
MPS simulations were performed using the ITensor library~\cite{itensor,itensor-r0.3}. 
\end{acknowledgments}


%

\end{document}



\title{Supplemental material: Charge-order melting in the one-dimensional Edwards model}

\author{Florian Lange}
\affiliation{Friedrich-Alexander-Universit\"at Erlangen-N\"urnberg (FAU), Erlangen National High Performance Computing Center (NHR@FAU), 91058 Erlangen, Germany}
\author{Gerhard Wellein}
\affiliation{Friedrich-Alexander-Universit\"at Erlangen-N\"urnberg (FAU), Erlangen National High Performance Computing Center (NHR@FAU), 91058 Erlangen, Germany}
\author{Holger Fehske}
\affiliation{Friedrich-Alexander-Universit\"at Erlangen-N\"urnberg (FAU), Erlangen National High Performance Computing Center (NHR@FAU), 91058 Erlangen, Germany}
\affiliation{Institute of Physics, University of Greifswald, 17489 Greifswald, Germany}

\date{\today}

\maketitle

\section{Effective $t$-$V$ Hamiltonian for the Edwards model}
Using standard methods (see, e.g., Ref.~\cite{PhysRevB.37.9753}), an effective Hamiltonian that approximates the Edwards model for large $\omega_0$ can be derived~\cite{PhysRevB.87.045116}. Taking $\omega_0 \sum_i \hat{b}_i^\dagger \hat{b}_i^{\phantom{\dagger}}$ as the unperturbed Hamiltonian, one obtains to second order 
\begin{align}
  \hat{H}^{(2)} = \omega_0 \sum_{j} \hat{b}_j^\dagger \hat{b}_j^{\phantom{\dagger}} - \frac{2 \lambda t_b}{\omega_0} \sum_{j} \left( \hat{f}_{j+1}^\dagger \hat{f}_{j}^{\phantom{\dagger}} + \text{H.c.} \right) &+ \frac{t_b^2}{\omega_0} \bigg[
    \sum_j \left( \hat{f}_{j+1}^\dagger \hat{f}_{j-1}^{\phantom{\dagger}}
    ( \hat{n}_j^f + \hat{n}_j^b - \hat{b}_{j-1}^\dagger \hat{b}_{j+1}^{\phantom{\dagger}} ) + \text{H.c.} \right) \nonumber \\
    &\hspace*{5mm} + 2 \sum_j \hat{n}_{j}^f \hat{n}_{j+1}^f- \sum_j \left(\hat{n}_{j+1}^f - \hat{n}_{j}^f \right)\left( \hat{n}_{j+1}^b - \hat{n}_j^b \right) 
    \bigg], 
  \label{eqPertHam}
\end{align}
where $\hat{n}_i^f = \hat{f}_i^\dagger \hat{f}_i^{\phantom{\dagger}}$ and $\hat{n}_i^b = \hat{b}_i^\dagger \hat{b}_i^{\phantom{\dagger}}$. For an empty boson system, this is the $t$-$V$ model for spinless fermions, except for an additional correlated next-nearest-neighbor hopping. 

\section{Detuned pumping}
\begin{figure}[tp]
  \centering
  \includegraphics[width=0.9\textwidth]{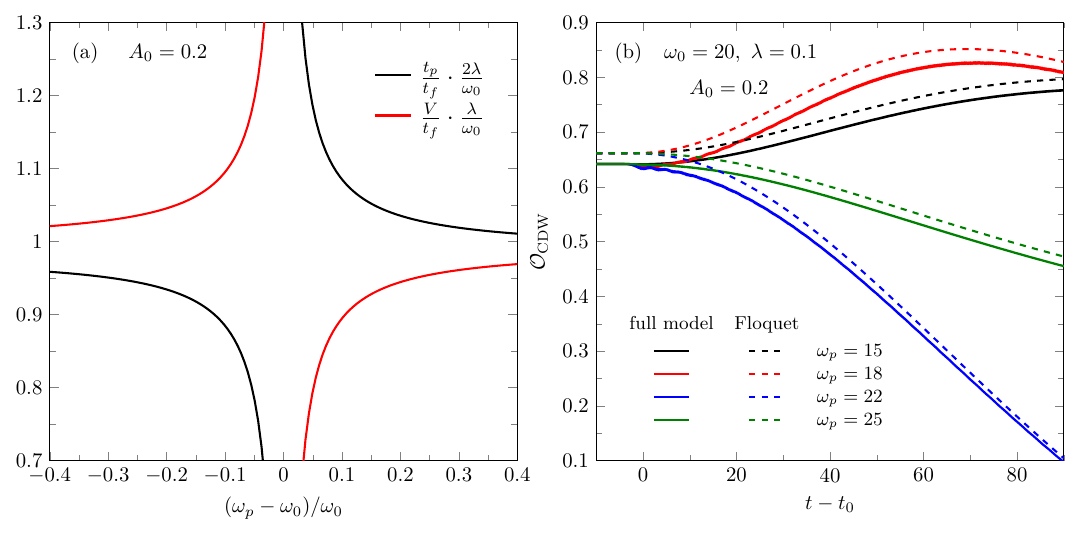}
  \caption{Dependence of the parameters $t_p$ and $V$ in the effective Floquet model~\eqref{eqHamFloquet} on the detuning of the pulse from the boson energy [panel (a)], and time evolution of the CDW order for a perturbation according to Eq.~\eqref{eqPertRamp} with $\sigma_p=2.0$ and $A_0=0.2$ [panel (b)]. The unit of energy is $t_b=1$.}
  \label{figSup}
\end{figure}

Let us now consider a system driven by a time-periodic vector potential $A(t) = A_0 \cos(\omega_p t)$. As described in Ref.~\cite{Claassen2017}, one can use Floquet theory in combination with a perturbative expansion to derive an effective Hamiltonian that captures the slow time-evolution for large boson energies and detuned driving frequency ($\omega_0, |\omega_p + m \omega_0| \gg t_b, \lambda ; \ \forall m \in \mathbb{Z}$). 
Assuming an empty boson background, we get to second order:
\begin{align}
  \hat{H}_{\text{Floquet}}^{(2)}(A_0,\omega_p) &= - t_f  \sum_j \left( \hat{f}_{j+1}^\dagger \hat{f}_{j}^{\phantom{\dagger}} + \text{H.c.} \right) + t_p \sum_j \left( \hat{f}_{j+1}^\dagger \hat{f}_{j-1}^{\phantom{\dagger}} \hat{n}_j^f + \text{H.c.} \right) + V \sum_j \hat{n}_{j}^f \hat{n}_{j+1}^f .
  \label{eqHamFloquet}
\end{align}
The parameters are
\begin{align}
  t_f &= \frac{2 \lambda t_b^{(0)}}{\omega_0} , \\
  t_p &= \sum_m \frac{t_b^{(-m)}t_b^{(m)}}{\omega_0 + m \omega_p} \\
  V &= 2 \sum_m \frac{\left|t_b^{(m)}\right|^2}{\omega_0 + m \omega_p},
  \label{eqEffV}
\end{align}
where
\begin{align}
  t_b^{(m)} &=  \frac{t_b}{2 \pi} \int_0^{2 \pi } d\tau \; e^{i m \tau} e^{i A_0 \cos(\tau) } . 
\end{align}
While the Hamiltonian $\hat{H}_{\text{Floquet}}^{(2)}(A_0,\omega_p)$ has the same form as $\hat{H}^{(2)}$ in Eq.~\eqref{eqPertHam} for zero bosons, the relative strengths of the terms are generally different when $A_0$ is finite [see Fig.~\ref{figSup} (a)]. 
In particular, Eq.~\eqref{eqEffV} indicates that for $\omega_p < \omega_0$ ($\omega_p > \omega_0$) the strength $V$ of the repulsive interaction increases (decreases) because of the term with $m = -1$. 
Note that the renormalized parameter $t_f$ depends on the amplitude $A_0$ but not the driving frequency $\omega_p$.

To test the validity of $\hat{H}_{\text{Floquet}}^{(2)}$, we simulate the time evolution of a system that starts from the ground state and is driven out of equilibrium by  a perturbation with the vector potential
\begin{align}
  A(t) &= A_0 e^{-\frac{t - t_0}{2 \sigma_p^2}\Theta(t_0-t) } \cos[\omega_p (t - t_0)],
  \label{eqPertRamp}
\end{align}
where $\Theta(t_0-t)$ is the step function. This initially is equal to a Gaussian pulse with width $\sigma_p$ and center $t_0$ as in the main text, but then remains at a constant amplitude $A_0$. In Fig.~\ref{figSup}~(b), we compare the direct simulation of the Edwards model with the simulation of the time-dependent effective model given by $\hat{H}(t) = \hat{H}_{\text{Floquet}}^{(2)}(A_0  e^{-\Theta(t_0-t)(t-t_0)/(2 \sigma_p^2)}, \omega_p) $. For sufficiently large $\omega_0$ and detuning, the Floquet model accurately predicts the evolution of the charge-density-wave (CDW) order $\mathcal{O}_{\text{CDW}}$. The main deviation is a shift due to the different CDW orders in the ground state, which decreases for larger $\omega_0$ (not shown).

